\documentclass[fleqn,usenatbib]{mnras}

\usepackage{newtxtext,newtxmath}
\usepackage[T1]{fontenc}

\DeclareRobustCommand{\VAN}[3]{#2}
\let\VANthebibliography\thebibliography
\def\thebibliography{\DeclareRobustCommand{\VAN}[3]{##3}\VANthebibliography}

\usepackage{graphicx}	% Including figure files
\usepackage{amsmath}	% Advanced maths commands
\usepackage{orcidlink}

\newcommand{\superstars}{\emph{Superstars}}

\title[Auriga Superstars]{Auriga \superstars: Improving the resolution and fidelity of stellar dynamics in cosmological galaxy simulations}

\author[R. Pakmor et al.]{R\"udiger Pakmor$^1$\thanks{rpakmor@mpa-garching.mpg.de}\orcidlink{0000-0003-3308-2420},
Francesca Fragkoudi$^{2}$\orcidlink{0000-0002-0897-3013},
Robert J.~J.~Grand$^{3}$\orcidlink{0000-0001-9667-1340},
Christine M. Simpson$^{4}$\orcidlink{0000-0001-9985-1814},
\newauthor
Facundo A.~G\'{o}mez,$^{5}$\orcidlink{0000-0003-4232-8584},
Freeke van~de~Voort$^6$\orcidlink{0000-0002-6301-638X},
Rebekka Bieri$^7$\orcidlink{0000-0002-4554-4488},
Wilma Trick$^1$\orcidlink{0000-0002-2806-3738},
\newauthor
Maria Werhahn$^1$\orcidlink{0000-0003-4984-4389}, 
Rosie Y. Talbot$^1$\orcidlink{0000-0001-9393-7879}
\vspace*{0.1cm}\\
$^{1}$Max-Planck-Institut f\"{u}r Astrophysik, Karl-Schwarzschild-Str. 1, D-85748, Garching, Germany\\
$^{2}$Institute for Computational Cosmology, Department of Physics, Durham University, South Road, Durham DH1 3LE, UK \\
$^{3}$Astrophysics Research Institute, Liverpool John Moores University, 146 Brownlow Hill, Liverpool, L3 5RF, UK\\
$^{4}$Argonne Leadership Computing Facility, Argonne National Laboratory, Lemont, IL 60439, USA\\
$^{5}$Departamento de Astronom\'{i}a, Universidad de La Serena, Av.~Juan Cisternas 1200 Norte, La Serena, Chile\\
$^{6}$Cardiff Hub for Astrophysics Research and Technology, School of Physics and Astronomy, Cardiff University, Queen’s Buildings, Cardiff CF24 3AA, UK\\
$^{7}$Department of Astrophysics, University of Zurich, 8057 Zurich, Switzerland\\
}

\date{Accepted 2025 October 9. Received 2025 October 9; in original form 2025 July 29}

\pubyear{2025}

\begin{document}
\label{firstpage}
\pagerange{\pageref{firstpage}--\pageref{lastpage}}
\maketitle

\begin{abstract}
Cosmological hydrodynamical simulations have become an indispensable tool to understand galaxies. However, computational constraints still severely limit their numerical resolution. This not only restricts the sampling of the stellar component and its direct comparison to detailed observations, but also the precision with which it is evolved. To overcome these problems we introduce the \superstars\ method. This method increases the stellar mass resolution in cosmological galaxy simulations in a computationally inexpensive way for a fixed dark matter and gas resolution without altering any global properties of the simulated galaxies. We demonstrate the \superstars\ method for a Milky Way-like galaxy of the Auriga project, improving the stellar mass resolution by factors of $8$ and $64$ at an additional cost of only $10\%$ and $500\%$, respectively. We show and quantify that this improves the sampling of the stellar population in the disc and halo without changing the properties of the central galaxy or its satellites, unlike simulations that change the resolution of all components (gas, dark matter, stars). Moreover, the better stellar mass resolution reduces numerical heating of the stellar disc in its outskirts and keeps substructures in the stellar disc and inner halo more coherent. It also makes lower mass and lower surface brightness structures in the stellar halo more visible. The \superstars\ method is straightforward to incorporate in any cosmological galaxy simulation that does not resolve individual stars.

\end{abstract}


%%%%%%%%%%%%%%%%%%%%%%%%%%%%%%%%%%%%%%%%%%%%%%%%%%

%%%%%%%%%%%%%%%%% BODY OF PAPER %%%%%%%%%%%%%%%%%%

\section{Introduction}

The Gaia satellite \citep{Gaia, GaiaDR1, GaiaDR2} and other recent observational campaigns have provided us with a detailed picture of the stellar discs and stellar halos of the Milky Way and nearby galaxies, in particular of various substructures in the disc and halo \citep[see, e.g., the reviews by][and references therein]{Helmi2020,Deason2024}. Stellar dynamics includes a multitude of complex phenomena, from the formation of bars and spiral arms in disc galaxies to tidal streams in the halo. Numerical simulations are a crucial tool to understand the non-linear physical processes that create and govern stellar structures. In particular, simulations that form and evolve galaxies in their full cosmological context are critical to understand stellar structures. Only cosmological simulations self-consistently include the formation and growth of galaxies over cosmic time, which is critical to study the long-lived structures like stellar bars and spiral arms and secular evolution in the stellar disc. In contrast to isolated setups, only cosmological simulations self-consistently include all interactions with satellite galaxies. These interactions with satellites shape the stellar halo \citep{Monachesi2019} and are a source of structures in stellar discs ranging from small-scale substructure \citep{Deason2024} to large-scale perturbations \citep{Gomez2016b,Grand2023}.
However, stellar substructures are sometimes small \citep{Helmi2020} and stellar streams \citep{Shipp2018} can consist of only thousands or even only hundreds of stars. These structures can also be old. In simulations, equivalent structures are therefore susceptible to numerical noise. We therefore need high-resolution cosmological galaxy simulations to model these structures at all. Specifically, we need high stellar mass resolution to accurately evolve stellar dynamics over long timescales, to be able to model and understand the formation and evolution of complex stellar structures. For example, simulating the formation of spiral arms might require more than $100$~million star particles in the stellar disc to not be dominated by noise \citep{DOnghia2013}.

\begin{figure*}
    \centering
    \includegraphics[width=0.95\textwidth]{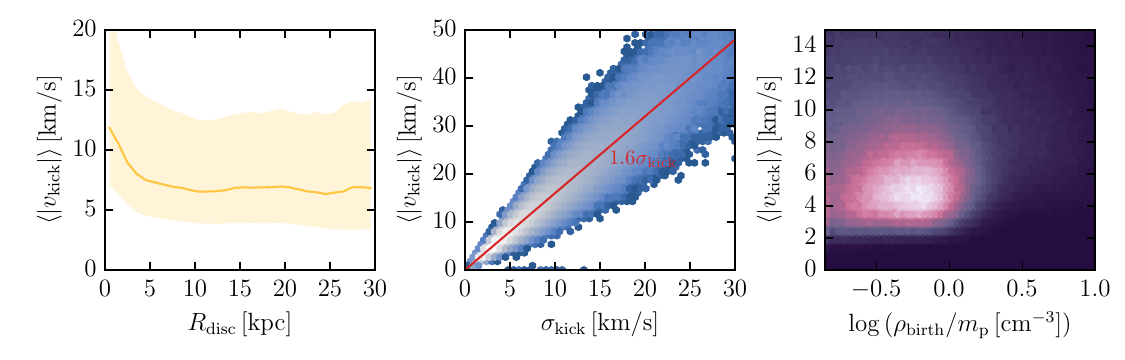}
    \caption{Quantification of the additional velocity kick that stars get at formation with the \superstars\ method. The panels show a simulation with $8$x better stellar mass resolution (Stars x8). The left panel shows the average absolute velocity kick imparted on star particles at formation as a function of their radial coordinate in the disc at $z=0$. The shaded area shows the $16$th and $84$th percentiles. We include all star particles with a vertical position $|z_\mathrm{disc}|<30~\mathrm{kpc}$ at $z=0$. The middle panel shows the average absolute velocity kick as a function of the input velocity dispersion of the birth cell. The right panel shows the average absolute velocity kick as a function of the density of the birth cell. All panels use mass-weighted averages for the Superstars particles. In the 2D histograms (middle and right panels) lighter colours correspond to more stars/higher total stellar mass per bin. Note the different range of values on the vertical axes. The typical additional velocity kick imparted on star particles at their birth is $5{-}10\,\mathrm{km/s}$.}
    \label{fig:creation}
\end{figure*}

In principle, cosmological simulations of Milky Way-like galaxies with exceptionally high mass resolution of $1000\,\rmn{M}_\odot$ per particle are possible \citep{Grand2021,Applebaum2021}. However, they are very computationally expensive and it is essentially impossible to properly calibrate the galaxy formation model for them, because the properties of the galaxies (most importantly their stellar mass) change too much with resolution \citep{Pillepich2018, FIRE2, Grand2021, Pathak2025}. For example, galaxies simulated with the Auriga galaxy formation model increase their stellar mass by $\sim30\%$ at $z=0$ for every factor of $8$ improvement in mass resolution of all components \citep{Auriga,Grand2021}. This change is very similar for galaxies simulated with the IllustrisTNG model \citep{Pillepich2018}. Consequently, increasing the mass resolution by only a factor of $8$ can significantly change the global and structural properties of a stellar disc or a satellite galaxy. For example, whether a central bar forms or not can differ with resolution \citep{Fragkoudi2025}. This limits our ability to draw conclusions about stellar structures in galaxy simulations that were run at better mass resolution than at which the model was calibrated\footnote{For example, the Auriga model was calibrated at a baryonic mass resolution of $5\times10^4\,\mathrm{M_\odot}$ \citep{Auriga}, the IllustrisTNG model was calibrated at a baryonic mass resolution of $10^6\,\mathrm{M_\odot}$ \citep{TNGMethodWeinberger,TNGMethodsPillepich}.}.

To sidestep this problem and to reach better stellar resolution without changing the galaxy significantly we introduce the \superstars\ method: Rather than forming a single massive star particle from a star-forming gas cell, we form several less massive star particles (each with a small additional velocity kick) in the same star formation event. This allows us to improve the mass resolution of the stellar component in galaxy simulations significantly, without changing the mass resolution of the gas, and without affecting the galactic wind model of Auriga or TNG. We show that this method is straightforward to include in any cosmological galaxy simulation where star particles represent average stellar populations. It improves not only the number of stellar tracers, but also the fidelity of the evolution of stellar structures, as it does not change the global properties of the stellar disc. Moreover, this method comes only at a moderate additional computational cost.

We explain the \superstars\ method in Section~\ref{sec:method}. We demonstrate in Section~\ref{sec:global} that it preserves the global properties of the galaxies, and notably of their stellar discs, very well and in particular significantly better than previous higher resolution Auriga simulations. In Section~\ref{sec:satellites} we focus on the satellite population and show that satellite properties remain unchanged with the \superstars\ method as well. In Section~\ref{sec:debris} we show how the \superstars\ method allows us to much better identify, characterise, and preserve substructures in the solar neighbourhood as well as structures in the stellar halo. We conclude with a summary and outlook in Section~\ref{sec:summary}.

\section{The Superstars method}
\label{sec:method}

We introduce the \superstars\ method on top of the Auriga galaxy formation model \citep{Auriga}, though it is generally applicable to any galaxy formation simulation. The Auriga galaxy formation model aims to simulate the formation and evolution of galaxies over cosmic history. It is implemented in the moving-mesh code \textsc{arepo} \citep{Arepo, Pakmor2016, Weinberger2020}, that solves the equations of magnetohydrodynamics on a moving Voronoi mesh with a second-order finite-volume scheme, fully coupled to self-gravity with a combined tree and particle-mesh solver. The Auriga galaxy formation model includes primordial and metal line cooling \citep{Vogelsberger2013}, an effective model for the interstellar medium and stochastic star formation \citep{Springel2003}, an effective model for galactic winds driven by stellar feedback, mass return from star particles via stellar winds and supernovae, and a model for the seeding, growth, and feedback from supermassive black holes \citep{Auriga}.

The basic idea of the \superstars\ method is to selectively improve the mass resolution of star particles, but keep the mass resolution of the gas the same. Ideally, this allows us to keep galaxy properties unchanged, that is, within the range of intrinsic variation of galaxy properties found among realisations run on a different machine or with a different random number seed. Such variations still happen, but are smaller than the systematic shifts caused by changing the mass resolution of the gas \citep{Genel2019,PakmorRealisations}.

The approach of the \superstars\ method to increase the stellar mass resolution begins with considering the stochastic formation of star particles within galaxy simulations. In general, cosmological galaxy simulations treat star formation stochastically. In the Auriga model, star-forming gas cells with a mass of $m_\mathrm{cell}$ and a non-zero star formation rate of $\dot{m}_\mathrm{SF}$ have a probability of $p_\mathrm{SF}$ to create a star particle of mass $m_*$ in a timestep $\Delta t_\mathrm{cell}$, given by
\begin{equation}
 p_\mathrm{SF} = \frac{m_\mathrm{cell}}{m_\mathrm{*}} \left( 1 - \exp{\left(- \frac{\dot{m}_\mathrm{SF} \, \Delta t_\mathrm{cell}}{m_\mathrm{cell}}\right)} \right)
\label{eq:sf}
\end{equation}
\citep{Springel2003}.

The default choice is to set the mass of the newly created star particle equal to the mass of the gas cell, that is $m_*=m_\mathrm{cell}$. If a cell is chosen for a star formation event, this typically leads to the conversion of the cell to a single star particle. One obvious way to improve the stellar mass resolution is to  create a single star particle of lower mass in the case of a star formation event, that is, to choose $m_* < m_\mathrm{cell}$. According to Equation~\ref{eq:sf}, this automatically increases the probability of creating a star particle in each timestep. The simulation would then form a larger number of less massive star particles. However, there is a practical limit to this approach, when the probability $p_\mathrm{SF}$ becomes larger than unity. In this case, the star formation rate can no longer be properly represented for the given timestep. We can circumvent this problem by introducing an additional timestep limiter, so that $P_\mathrm{SF}<1$. Because this timestep criterion scales linearly with the stellar mass resolution, it quickly becomes dominant for star-forming cells over the classic Courant-Friedrichs-Lewy timestep criterion $\Delta t < R_\mathrm{cell} / c_\mathrm{s,cell}$. This makes the approach to just set $m_*<m_\mathrm{cell}$ and create single less massive star particles very expensive and in practice unfeasible for achieving significantly better stellar mass resolution. Nevertheless, this method has been used to improve the stellar mass resolution by a factor of a few \citep{Applebaum2021}.

\begin{table}
\begin{center}
\begin{tabular}{ l r r r }
\hline
Name & $m_\rmn{DM}$ & $m_\rmn{gas}$ & $m_\rmn{*}$ \\
& $[10^3\rmn{M}_\odot]$ & $[10^3\rmn{M}_\odot]$ & $[10^3\rmn{M}_\odot]$\\
\hline
Reference\,(L4;\,Stars\,x1) & $300$ & $50$ & $50$\\
\hline
Stars\,x8 & $300$ & $50$ & $6$\\
\hline
Stars\,x64 & $300$ & $50$ & $0.8$\\
\hline
DM\,x8 & $40$ & $50$ & $50$\\
\hline
$\mathrm{All\,x8\,(L3)}$ & $40$ & $6$ & $6$\\
\hline
$\mathrm{All\,x64\,(L2)}$ & $5$ & $0.8$ & $0.8$\\
\hline
\hline
\end{tabular}
\end{center}
\caption{Mass resolution of dark matter particles, gas cells, and star particles in the simulations shown in this paper. Each row corresponds to one simulation setup. For Reference (L4, Stars x1), Stars x8 and DM x8 we run more than one realisation \citep{PakmorRealisations}.}
\label{tab:resolution}
\end{table}

\begin{table*}
\begin{center}
\setlength\tabcolsep{5pt}
\begin{tabular}{ l c c c c c c c c c c c }
\hline
Name & $N_\mathrm{sims}$ & $M_{200\mathrm{c}}$ & $M_*$ & $M_{*,r>50\mathrm{kpc}}$ & $\dot{M}_{*,1\mathrm{Gyr}}$ & $V_\mathrm{c,max}$ & $R_\mathrm{opt}$ & $R_\mathrm{disc}$ & $H_\mathrm{disc}$ & $\sigma_{*,\mathrm{r}}$ & $\sigma_{*,\mathrm{z}}$ \\
& & $[10^{12}\rmn{M}_\odot]$ & $[10^{10}\rmn{M}_\odot]$ & $[10^{10}\rmn{M}_\odot]$ & $[\rmn{M}_\odot / \rmn{yr}]$ & $[\rmn{km/s}]$ & $[\rmn{km/s}]$ & $[\rmn{kpc}]$ & $[\rmn{kpc}]$ & $[\rmn{km/s}]$ & $[\rmn{km/s}]$\\
\hline
Reference\,(L4;\,Stars\,x1) & 7 & $1.04\pm0.01$ & $5.4\pm0.2$ & $0.42\pm0.02$ & $2.4\pm0.8$ & $ 207\pm   5$ & $14.1\pm 1.3$& $4.5\pm0.5$ & $0.89\pm0.03$ & $48\pm 2$ & $39\pm 1$\\
\hline
Stars\,x8 & 7 & $1.03\pm0.01$ & $5.4\pm0.3$ & $0.41\pm0.01$ & $1.9\pm0.7$ & $ 205\pm   4$ & $12.8\pm 1.7$& $4.9\pm0.4$ & $0.90\pm0.08$ & $47\pm 1$ & $39\pm 1$\\
\hline
Stars\,x64 & 1 & $1.03$ & $5.3$ & $0.40$ & $1.8$ & $ 203$ & $14.3$ & $5.3$ & $0.99$ & $49$ & $39$\\
\hline
DM\,x8 & 3 & $1.03\pm0.00$ & $5.5\pm0.1$ & $0.45\pm0.01$ & $2.7\pm1.2$ & $ 205\pm   1$ & $15.2\pm 3.2$& $4.6\pm0.1$ & $0.97\pm0.02$ & $49\pm 0$ & $41\pm 1$\\
\hline
$\mathrm{All\,x8\,(L3)}$ & 1 & $1.01$ & $6.6$ & $0.53$ & $2.0$ & $ 237$ & $12.2$ & $3.8$ & $1.04$ & $52$ & $44$\\
\hline
$\mathrm{All\,x64\,(L2)}$ & 1 & $1.02$ & $7.6$ & $0.80$ & $3.1$ & $ 225$ & $13.3$ & $3.5$ & $1.96$ & $69$ & $56$\\
\hline
\hline
\end{tabular}
\end{center}
\caption{Global properties of the simulated galaxies at $z=0$. They all simulate halo Au-6 and only change the numerical resolution of different components of the simulation as summarised in Table~\ref{tab:resolution}. The rows show the reference (L4, standard resolution for all components) simulations, the simulations with $8\times$ and $64\times$ better mass resolution for the star particles, the simulations with $8$x better mass resolution for dark matter, and the simulations with $8\times$ (L3) and $64$x (L2) better mass resolution for all components. $N_\mathrm{sims}$ is the number of realisations we have run a setup in. If $N_\mathrm{sims}>1$ we quote the mean value and the variance over all $N_\mathrm{sims}$ simulations of this type. The other columns show the halo mass, $M_{200\mathrm{c}}$, the total stellar mass of the central galaxy, $M_*$, the stellar mass in the halo outside $50\,\mathrm{kpc}$ including satellite galaxies, $M_{*,r>50\mathrm{kpc}}$, the stellar mass formed in the central galaxy in the last Gyr, $\dot{M}_{*,1\mathrm{Gyr}}$, the maximum of the rotation curve, $V_\mathrm{c,max}$, the optical radius, $R_\mathrm{opt}$, defined as the radius at which the surface brightness in the $B$-band drops below $25\,\mathrm{mag}\,\mathrm{arcsec}^{-2}$, the scale radius, $R_\mathrm{disc}$, and scale height, $H_\mathrm{disc}$, of the stellar disc, and the radial velocity dispersion, $\sigma_{*,\mathrm{r}}$, and vertical velocity dispersion, $\sigma_{*,\mathrm{z}}$, at a radius of $8\,\mathrm{kpc}$ in the disc. Almost all quantities remain consistent with the reference run when changing stellar or dark matter mass resolution only, but change significacntly when changing the gas mass resolution as well.}
\label{tab:simulations}
\end{table*}

We therefore take a different approach that is closer to the original method that typically converts the whole mass of a cell into one star particle. We still set $m_*=m_\mathrm{cell}$ in Equation~\ref{eq:sf}. Then, when a cell is converted to stars, we create $N$ star particles with a total mass of $m_\mathrm{cell}$ at the same time rather than only one star particle. Each star particle then receives a mass of $m_\mathrm{cell}/N$. All star particles created from a single cell inherit the position and velocity of the cell.

However, we do not want all star particles created together to move on exactly identical trajectories, because even though we have more particles it would not add any additional information to the simulation over evolving a single more massive star particle. Therefore, we add a small additional velocity kick to each newly born star particle, where each kick is unique to the particle. This can, for example, be interpreted as the velocity dispersion in a birth cloud that formed those stars, or just as an additional parameter of the star formation model. We require this additional velocity to be large enough such that star particles born together are not bound to each other, and small enough to not change any properties of the galaxy.

If a cell is selected to form stars in the \superstars\ method, we first estimate the magnitude of the kick we want to impart on the new star particles that are about to be created from this cell. We set it to the minimum of the gas velocity dispersion of the cell and its direct neighbours, limited by the sound speed of the cell, that is,
\begin{equation}
   \sigma_\mathrm{kick} = \mathrm{min}\left(\sigma_\mathrm{cell}, c_\mathrm{s,cell}\right) .
\end{equation}

This can be seen as a measurement of the magnitude of unresolved velocities on scales just below the grid scale. This choice, however, is relatively arbitrary. In principle, we could also choose a constant value of, for example, $10\,\mathrm{km/s}$ as a typical value of small-scale velocity fluctuations in the interstellar medium below our resolution scale. However, here we attempt to include some of the expected variations in the properties of the interstellar medium. Note that in the Auriga model the sound speed in star-forming cells is set by an effective equation of state for the interstellar medium \citep{Springel2003}. In this picture, a higher sound speed of star-forming cells represents an unresolved turbulent velocity field with higher velocities. This motivates why we allow for larger birth velocity kicks in higher density environments. Including this velocity might also avoid artificially very cool stellar discs in isolated galaxies with a very quiet interstellar medium \citep{Burger2025}. One obvious extension in future work is to couple the kick velocity to a fully fledged turbulence subgrid model \citep[see, for example][]{Semenov2016,Kretschmer2020}. This would also allow us to estimate the turbulent velocity dispersion on an appropriate physical scale, rather than only on the scale of the local cell size.

For each newly created star particle we then compute an individual kick velocity. We draw each component of the kick velocity vector from a normal distribution with a mean of zero and a width of $\sigma_\mathrm{kick}$, that is,
\begin{equation}
   v_\mathrm{kick,i} = \mathcal{N} \left( 0, \sigma_\mathrm{kick} \right) .
\end{equation}

The absolute value of the kick velocity then follows a Chi distribution with three degrees of freedom, so its expectation value will scale with $\sigma_\mathrm{kick}$ as
\begin{equation}
     \langle \left| \vec{v}_\mathrm{kick} \right| \rangle = \sigma_\mathrm{kick} \cdot \sqrt{\frac{8}{\pi}} \approx 1.6 \cdot \sigma_\mathrm{kick} .
     \label{ref:chi}
\end{equation}

The full initial velocity of a newly born particle then becomes
\begin{equation}
   \vec{v}_\mathrm{*} = \vec{v}_\mathrm{cell} + \vec{v}_\mathrm{kick} .
\end{equation}

Finally, we subtract the mass-weighted average kick velocity vector from all star particles born together, so that the total added momentum from the kicks is zero. We do not change any aspect of the wind particles that model feedback on galactic scales \citep{Auriga}. This means that wind particles are still formed by converting a cell into a single wind particle of the same mass. We compute mass return from star particles into the interstellar medium for every star particle individually.

The left panel of Figure~\ref{fig:creation} shows the average absolute velocity of the birth kick as a function of the cylindrical radius of a star particle in the galactic disc at $z=0$ of a Milky Way-like galaxy at the standard gas resolution ($5\times10^4\,\mathrm{M_\odot}$) with $8\times$ better stellar mass resolution. That is, we form $8$ star particles in every star formation event. The distribution is essentially flat in radius with a median of about $7\,\mathrm{km/s}$. Only in the centre of the disc at $R_\mathrm{disc} \lesssim 3\,\mathrm{kpc}$ it increases slightly above $10\,\mathrm{km/s}$, likely as a result of stronger turbulence there. In the middle panel of Figure~\ref{fig:creation} we show that average absolute velocity of the birth kick is consistent with the expectation value of $1.6 \cdot \sigma_\mathrm{kick}$ (see Equation~\ref{ref:chi}). Finally, in the right panel of Figure~\ref{fig:creation} we show the average absolute velocity of the birth kick of star particles as a function of their birth density, that is, the density of the gas cell at the time when it made the star particles. Most stars are formed at birth densities just below $1\,\mathrm{cm^{-3}}/m_\mathrm{p}$ and the distribution of birth velocity kicks peaks around $5\,\mathrm{km/s}$ with a long tail to larger velocities. The tail is a result of turbulent gas with high velocity dispersion at the time of star formation.

We can estimate the escape velocity for a cloud of particles born together from their total mass ($5\times10^4\mathrm{M_\odot}$, the gas mass resolution) and the stellar softening ($250\,\mathrm{pc}$) to be roughly $1\,\mathrm{km/s}$. Therefore, star particles born together should essentially never be gravitationally bound to each other. For the simulations in this work we decided to always keep the softening of dark matter and stars the same to minimise numerical heating. Thus, in the simulations in which we improve the stellar mass resolution, but keep the dark matter mass resolution the same, we keep the standard softening for stars ($370\,\mathrm{pc}$). In the simulation with improved dark matter mass resolution we decrease the softening of both dark matter and star particles to $125\,\mathrm{pc}$, even when we do not improve the stellar mass resolution.

We only inject mass and metals from stars into the cell they are currently in. This significantly improves computational efficiency compared to injecting into the closest $64$ cells as originally done in the Auriga and IllustrisTNG models \citep{Auriga}. This also reduces over-mixing of metals at injection, because supernovae typically mix with $\sim10^4\,\mathrm{M_\odot}$ before their remnant disappears, which is equal or smaller than the mass of a single cell in our simulations. Single cell injection otherwise does not change the gas metallicity and stellar abundances much \citep{Vandevoort2020,Vandevoort2022}.

In the rest of this paper, we will first establish that the \superstars\ method with a typical additional birth velocity dispersion of $5-10\,\mathrm{km/s}$ does not change any global or dynamical properties of the galactic disc for a Milky Way-like galaxy in a cosmological simulation. Thereafter, we will show that the better stellar mass resolution not only carries additional information in the simulation, but also improves the fidelity of the modelling of the stellar component. This will allow us to better study stellar properties and dynamics in cosmological galaxy simulations in future work. As a rule of thumb, we see that a simulation with $8\times$ better stellar mass resolution is only about $10\%$ more expensive than the standard resolution. At this resolution the star particles are still outnumbered by the dark matter particles. Our simulation with $64\times$ better stellar mass resolution is about $6\times$ more expensive than the standard simulation. At this resolution the collisionless particles are dominated by the star particles. In comparison, a simulation with $8\times$ better mass resolution in all components increases the computational cost typically by a factor of $20$, so a simulation with $64\times$ better mass resolution in all components is $400\times$ more expensive.

\begin{figure*}
    \centering
    \includegraphics[width=\textwidth]{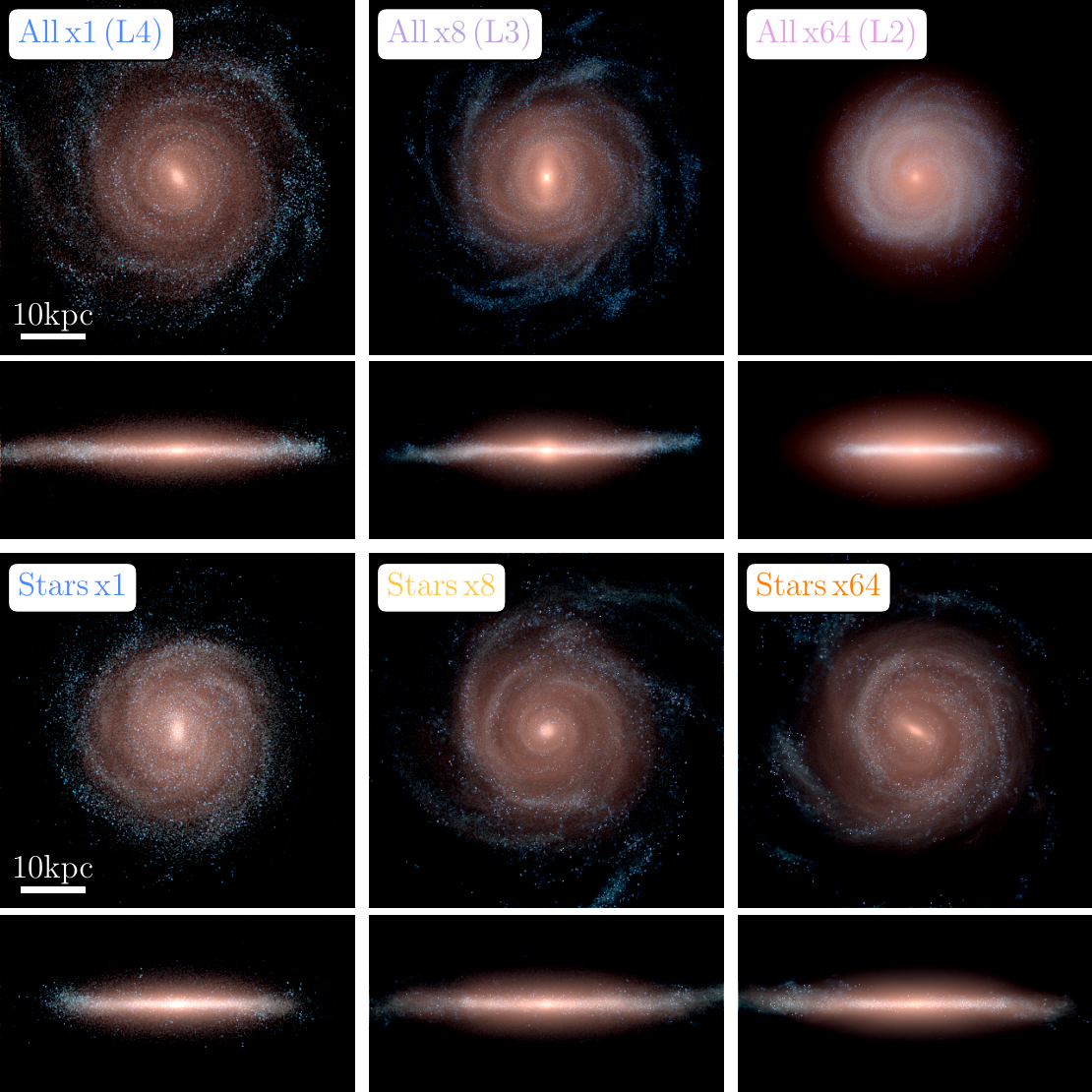}
    \caption{Face-on and edge-on stellar light projections at $z=0$ for three different resolution runs of the Auriga project (top rows) and for three runs increasing the stellar mass resolution only. The images show the $K$-band, $B$-band, and $U$-band as RGB channels. The left column shows the galaxy at the exact same resolution, but for a run with a different random number seed. Changing the resolution of all components (gas, dark matter, stars) clearly changes the global properties of the galaxy in a systematic way, but this is not the case when we only increase the stellar mass resolution with the \superstars\ method.}
    \label{fig:stellarlight}
\end{figure*}

\begin{figure*}
    \centering
    \includegraphics[width=\textwidth]{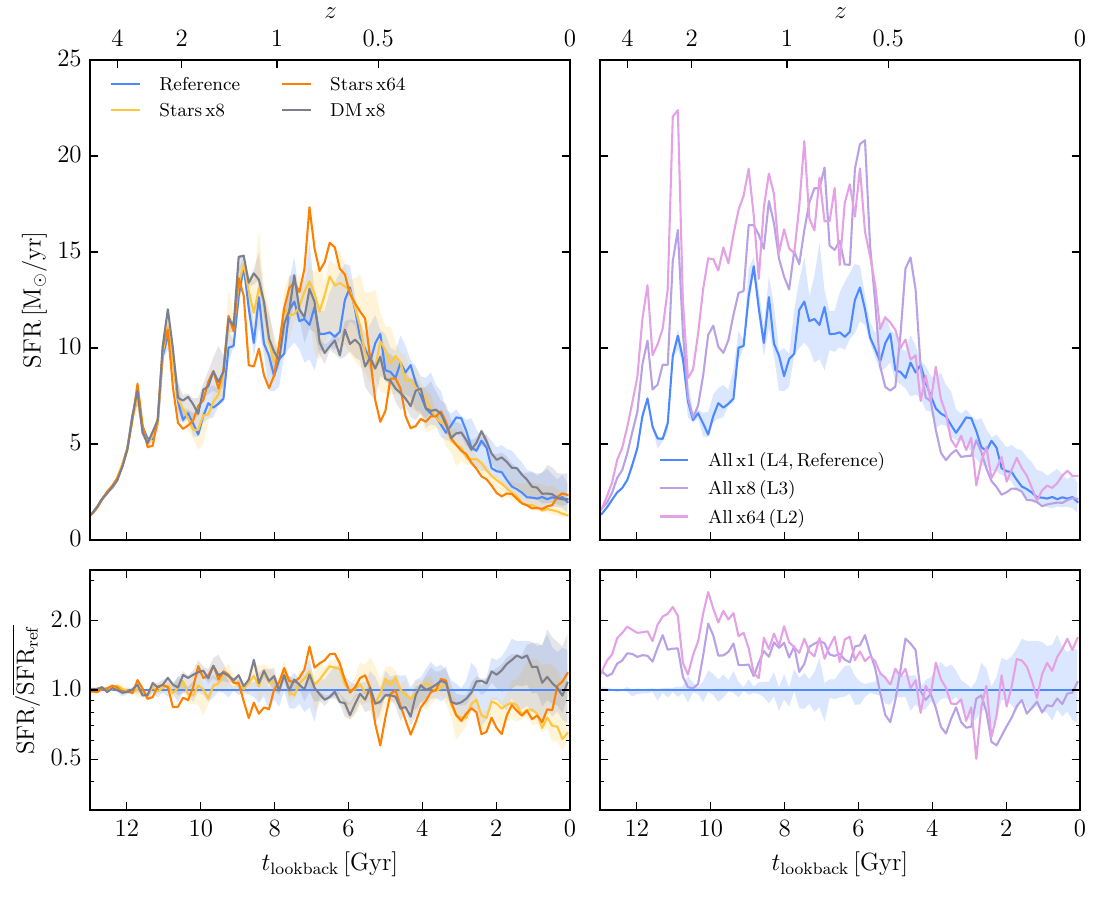}
    \caption{Star formation history of simulations that only change the mass resolution of one component, stellar mass or dark matter, (left panels) and for simulations that change the resolution of all components, that is, gas, stars, and dark matter (right panel). The lines show the median and the shaded bands the $16$th-$84$th percentiles of all realisations run for the different configurations (see Table~\ref{tab:simulations}; note that for the computationally expensive setups "Stars $\times64$" (orange), "All $\times8$ (L3)" (purple) and "All $\times64$ (L2)" (pink) only one realisation is available and therefore only a single line is shown). The star formation history remains the same if we change only the mass resolution of the dark matter or only the stars. In contrast, the star formation rate is significantly and systematically increased at times before $z=0.5$ if we change the gas resolution as well.}
    \label{fig:sfh}
\end{figure*}

\begin{figure*}
    \centering
    \includegraphics[width=\textwidth]{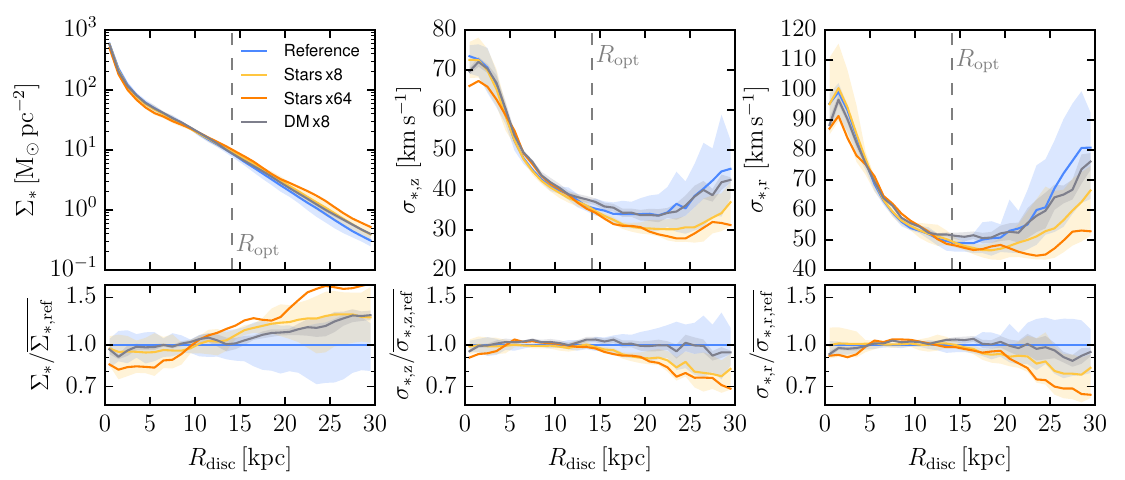}
    \caption{Comparison of the properties of the galactic stellar discs at $z=0$ for realisations with different stellar mass and dark matter mass resolutions, but the same gas mass resolution. The columns show the face-on stellar mass surface density computed with a depth $|z|<30\,\mathrm{kpc}$ (left panel), the vertical velocity dispersion (middle panel), and the radial velocity dispersion of the stars in the disc. The lines show the median and the shaded bands the $16$th-$84$th percentiles of all realisations run for the different configurations (see Table~\ref{tab:simulations}). The vertical dashed gray lines show the average optical radius of the reference simulations (blue). The bottom row shows the deviation from the median of the reference simulations. Neither increasing the stellar mass resolution, nor the dark matter mass resolution changes any properties of the stellar disc significantly for the runs with $8\times$ better stellar mass or dark matter mass resolution. At large radii $R_\mathrm{disc}>20\,\mathrm{kpc}$ the simulation with the $64$x better stellar mass resolution (orange) shows an increases in the surface density and a decrease in the velocity dispersions. However, it is unclear from this single, computationally expensive realisation how systematic this difference is.}
    \label{fig:stars}
\end{figure*}

\section{Global galaxy properties with the Superstars method}
\label{sec:global}

To test the method, we rerun Au-6 of the original Auriga suite \citep{Auriga}, which is similar to the Milky Way in stellar mass and properties of the stellar disc at $z=0$ and in its relatively quiet accretion history. We use the standard Auriga resolution (L4), that uses a baryonic mass resolution of $5\times10^4\,\mathrm{M_\odot}$, a dark matter mass resolution of $3\times10^5\,\mathrm{M_\odot}$, and a softening of $250\,\mathrm{pc}$ for dark matter and star particles, as our reference. The Auriga model was originally developed for this resolution.

We perform two resimulations of Au-6 with the \superstars\ method: one with $8\times$ better mass resolution for the stars; and one with $64\times$ better mass resolution for the stars. In both cases, we keep the mass resolution of dark matter and gas unchanged. For comparison, and to disentangle different numerical effects, we also resimulate Au-6 with $8\times$ better dark matter mass resolution only, but unchanged gas and stellar mass resolution. For all these simulations, we use the same gravitational softening lengths as in the original Auriga simulation at the reference resolution, irrespective of stellar and dark matter mass resolution. We also compare our results to the standard higher resolution simulations of Au-6 of the Auriga project, which increase the mass resolution of all components (gas, dark matter, stars) by fixed factors of $8$ (L3) and $64$ (L2). We show an overview of the mass resolution used for the different components for all simulations we use in this paper in Table~\ref{tab:resolution}.

We summarise these simulations and various properties of the main galaxy at $z=0$ in Table~\ref{tab:simulations}. To separate which differences are caused by variance within the model, and which differences are systematically caused by changes in the resolution, we ran $7$ realisations of the simulations with $8\times$ better stellar mass resolution with the \superstars\ method and $3$ realisations of the simulations with $8\times$ better dark matter mass resolution. We also include all $7$ realisations of the original Au-6 galaxy at L4 resolution discussed in detail in \citet{PakmorRealisations}. Here, \textit{realisations} refers to rerunning the simulation with exactly the same setup, initial conditions, and parameters, but changing the random number seed that in particular realises the stochastic star formation. All realisations of the same simulation are equally valid, so their small but non-negligible spread tells us about the intrinsic uncertainty of the model and simulation \citep{PakmorRealisations}. Due to the computational expense we only ran one realisation of the simulation with $64\times$ better stellar mass resolution.

For a first qualitative visual impression, we show stellar light projections of different versions of the same halo at $z=0$ in Figure~\ref{fig:stellarlight}. The left column shows two different realisations at Auriga's reference resolution. The top row then increases the resolution of all components by factors of $8$ (middle column, L3) and $64$ (right column, L2). We see, most notably in the old (red) component of the edge-on projections, that more stellar mass is formed at higher resolution, and that the size of the disc changes noticeably. For the highest resolution simulation the height of the disc also increases significantly. The latter is a result of a higher stellar mass of the simulation and a partial failure of the repositioning scheme for the central supermassive black hole in the main galaxy. This scheme is supposed to keep the black hole in the centre of the galaxy, compensating for the lack of dynamical friction because the mass of the black hole and the dark matter particles in the simulation are too close. It failed in the particular implementation of re-centering because the estimate of the local potential in a fixed mass of gas around the black hole at this high resolution included many gas cells and in particular local minima that prevented the black hole from always being repositioned towards the centre of the global potential of the galaxy. Its failure allowed the black hole to wander around in the galaxy several kpc from the centre rather than keeping it in the centre, significantly heating the stellar disc \citep{Grand2021}. The problem fundamentally only appears at the L2 resolution. It is therefore essentially impossible to find this problem in test runs, short of already running the full expensive cosmological zoom simulation.

In the bottom row of Figure~\ref{fig:stellarlight} we only increase the stellar mass resolution of the simulations with the \superstars\ method. The middle panel shows a simulation with $8\times$ better stellar mass resolution, and the right panel a simulation with $64\times$ better stellar mass resolution, but the same mass resolution for dark matter and gas as the reference run. The stellar discs look very similar, in particular in their old (red) stellar component, size, and height. The differences in the young stellar component are still in the range of intrinsic variance of the model \citep{PakmorRealisations}, that manifests for example as the difference between the upper left and lower left panels.

Moreover, the simulations with better stellar mass resolution result in sharper structures within the disc. We would expect the latter for the standard higher resolution Auriga runs in the top row as well, though this is counteracted for the highest resolution L2 run by the numerical problem with the black hole repositioning. Moreover, the higher total mass at higher resolution in the standard Auriga runs makes the disc more stable.

In Figure~\ref{fig:sfh} we show star formation histories of the halo without satellite galaxies (top panels) and their relative deviation from the reference simulations (bottom panels). The star formation histories shown in the left column are completely consistent with each other. These simulations change the stellar mass resolution or dark matter mass resolution, but keep the gas mass resolution the same. We therefore conclude that, as long as the gas resolution is kept constant, the full star formation history of the galaxy remains the same within variance with the \superstars\ method. In contrast, when we improve the gas mass resolution as well, as shown in the right panel of Figure~\ref{fig:sfh}, the star formation rate increases systematically at all times before $z\sim0.5$, leading to a significantly higher total stellar mass in these galaxies, as also shown in Table~\ref{tab:simulations} and visible in Figure~\ref{fig:stellarlight}. This fundamentally changes all properties of the galaxy, and essentially invalidates the model calibration at a resolution different than the standard resolution used for calibration of $m_\mathrm{gas}=5\times10^{4}\,\mathrm{M_\odot}$ for the Auriga model. Future efforts to run cosmological zoom simulations with better gas mass resolution would have to change the model so this change in stellar mass does not happen. However, recalibrating the model with better gas mass resolution would be computationally very expensive because of the large number of realisations required and therefore likely unfeasible.

In the following, we will focus on the differences introduced by the \superstars\ method and by changing only the dark matter mass resolution, and no longer consider the simulations that also change the gas mass \citep{Grand2021}. We show radial profiles of the stellar mass surface density of the galactic discs at $z=0$ and profiles of the vertical and radial velocity dispersions in Figure~\ref{fig:stars}. The lines represent the mean of all realisations at each radius, and the coloured bands show the variance across realisations. Here, we compute the variance of a distribution as the sample variance including Bessel's correction. We clearly see that for all $3$ profiles all simulations shown are largely consistent with each other within the optical radius. This result is particularly interesting for the profiles of the velocity dispersions, because we add an additional dispersion to newly born stars in the simulations with better stellar mass resolution. Following Figure~\ref{fig:stars} we argue that this kick is small enough to not change the velocity dispersion of the global stellar population in the disc, which is instead set by properties of the star-forming interstellar medium and galactic dynamics. We explicitly show the values of the velocity dispersions measured at the solar circle ($R_\mathrm{cyl}=8\,\mathrm{kpc}$) in Table~\ref{tab:simulations}.

There seems to be a systematic trend that the stellar surface density increases at larger radii and decreases slightly in the centre in simulations with better stellar mass resolution. This redistribution of stellar mass already happens at formation, that is, the stars are born at different radii (not shown). We attribute this to different dynamics and structures in the disc. For example, there are more pronounced spiral arms in the simulations with better stellar mass resolution (see also Figure~\ref{fig:stellarlight}) that redistribute the gas in the disc before it forms stars. Nevertheless, the total stellar mass in the disc remains the same (see Table~\ref{tab:simulations}). The more pronounced spiral arms in the simulations with better stellar mass could be a result of less numerical noise in the stellar potential that tends to wash out structures. We show a similar effect for substructures in Section~\ref{sec:debris}. We will investigate these differences, in particular in the context of spiral arms, in detail in future studies.

It is interesting to note that the simulations with higher dark matter mass resolution have a very similar velocity dispersion profile as the original simulations. In contrast, the simulations with better stellar mass resolution show a slightly smaller velocity dispersion at large radii ($R_\mathrm{cyl}\gtrsim15\,\mathrm{kpc}$). The simulations with $8\times$ better stellar mass resolution still overlap within the variance, but the single simulation with $64\times$ better stellar mass resolution does not. We argue that this is most easily explained by noise in the stellar disc potential at large radii. Because the stellar surface density is much lower in the outskirts of the disc than in its inner parts, the number of star particles representing the stellar disc potential is low, and noise in the potential might increase the stellar velocity dispersion at lower resolution. With $64\times$ better stellar mass resolution, the velocity dispersion profile becomes essentially constant at large radii indicating that the increase in the velocity dispersion at large radii seen in the simulations at the standard Auriga resolution might be entirely numerical. We will investigate this in more detail in future work.

The consistency between the velocity dispersion profiles of the reference simulations and the simulations with better dark matter resolution is also interesting in the context of spurious numerical heating driven by differences between the mass of dark matter and star particles \citep{Ludlow2021,Ludlow2023}. When there is a larger difference between these two types of collisionless particles, more massive particles will impart additional kinetic energy to less massive particles as the system evolves toward equipartition. However, the timescale on which this heating operates also depends on the total number of particles. Higher particle numbers mean that the particles are more collisionless, and that the timescale becomes longer. From analytical estimates, we do not expect the Auriga simulations or our \superstars\ variants to be affected by this problem for Milky Way-mass galaxies, because the dark matter mass resolution of the reference simulations is already sufficiently high such that spurious heating happens on timescales longer than the Hubble time. The identical velocity dispersion profile of the reference runs and the simulations with $8\times$ better dark matter mass resolution shown in Figure~\ref{fig:stars} demonstrates this explicitly.

Another interesting aspect is that the additional kicks to the star particles at birth do not change the overall velocity dispersion. In contrast, in idealised non-cosmological galaxy simulations with a similar effective model for the interstellar medium, the lack of turbulence in the disc can create a very cold stellar disc \citep{Burger2025}. In cosmological simulations, however, there is a significant amount of turbulence in the star-forming gas, despite the lack of explicit supernova feedback. This turbulence is likely driven by a combination of inflows onto the disc, differential rotation and shear instabilities in the disc, and instabilities in the shear layer between the disc and the circumgalactic medium \citep{PfrommerMonsterPaper}. Therefore, while potentially critical in non-cosmological galaxy simulations to obtain more realistic stellar discs, the introduction of an additional velocity dispersion when stars are born in cosmological simulations is not necessary. It is required by construction though in methods where the stellar mass resolution is increased using a similar approach to \superstars.

We present various other global quantities of the galaxies at $z=0$ in Table~\ref{tab:simulations} to show that they are not significantly modified by the \superstars\ method. In particular we show the stellar mass of the halo (defined as stellar mass between $R=50\,\mathrm{kpc}$ and $R_\mathrm{200c}$, including satellite galaxies), the stellar mass formed within the last Gyr, the maximum circular velocity, the scale radius and scale height of the stellar disc obtained from an exponential fit to the surface density, and the radial and vertical velocity dispersion at a radius of $8$~kpc. All the properties show negligible variation for changes in stellar mass resolution with the \superstars\ method or changes in the dark matter mass resolution. Notably, the stellar mass in the halo increases by about $10\%$ in the simulations with better dark matter mass resolution, because more satellite galaxies form at the low mass end \citep{Grand2021}. The scale radius and height of the galaxy in the simulation with $64\times$ better stellar mass resolution increase slightly, but it is unclear whether this effect is systematic or just variance.

In contrast, we see larger differences between the properties of the galaxies in the simulations where the gas mass resolution is increased alongside that of the stars and dark matter. With better mass resolution they slightly, but systematically, increase the stellar mass of the central galaxy, the stellar mass in the halo, and the maximum circular velocity. In addition, the discs become more compact and thicker. However, the significant change in stellar mass already alters the properties of the disc sufficiently that its internal structure and dynamics will no longer be comparable. This problem is clearly absent with the \superstars\ method and the galaxy properties essentially remain within the intrinsic variance of the model.

\begin{figure}
    \centering
    \includegraphics[width=\linewidth]{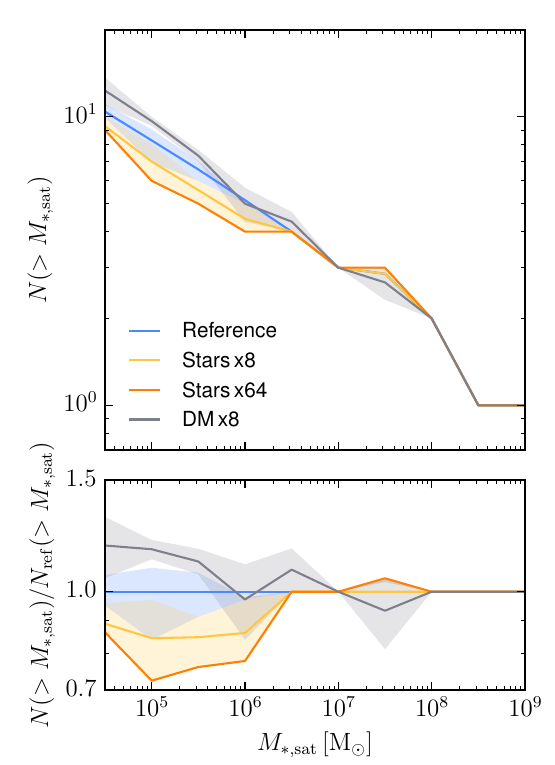}
    \caption{Satellite stellar mass function at $z=0$ for the reference run and simulations with improved stellar mass and dark matter mass resolution. The lower panel shows the satellite stellar mass function relative to the median of the reference runs. For a satellite stellar mass $M_{*,sat}>10^6\,\mathrm{M_\odot}$ the mass function is identical. Only for lower satellite stellar masses is there an indication that there are systematically slightly more satellites in the simulations with higher dark matter resolution (grey) and slightly fewer satellites in the simulations with higher stellar mass resolution (yellow, orange).}
    \label{fig:satellites}
\end{figure}

\begin{figure}
    \centering
    \includegraphics[width=\linewidth]{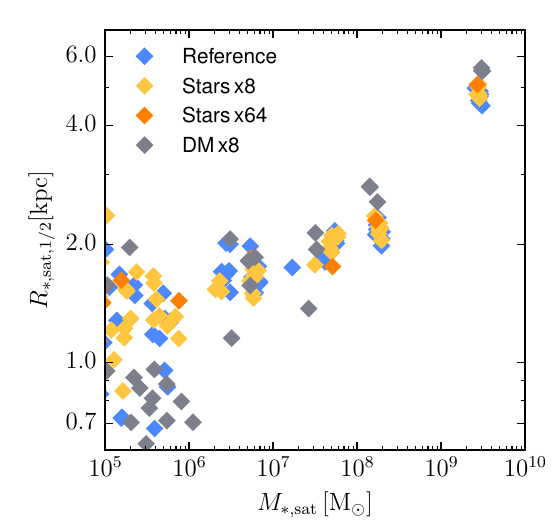}
    \caption{Stellar half mass radius of satellites at $z=0$ as a function of their stellar mass for the reference run and simulations with improved stellar mass and dark matter mass resolution. The stellar half mass radii are consistent between different simulations. In the simulations with better dark matter resolution the lowest mass satellites can be smaller (while being also more plentiful; see Figure \ref{fig:satellites}).}
    \label{fig:satellitesmassradius}
\end{figure}

\begin{figure}
    \centering
    \includegraphics[width=\linewidth]{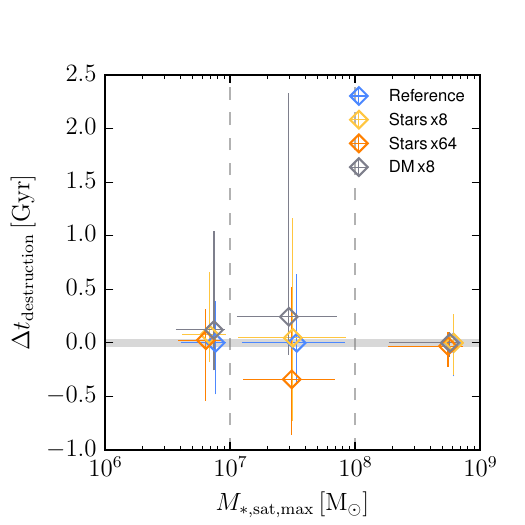}
    \caption{Time when satellites of different peak stellar mass ($M_{*,\rmn{sat,max}}$) were disrupted relative to the median time when the same satellites were disrupted in the reference runs. The data is binned in three stellar mass bins, covering a decade in stellar mass each (separated by vertical dashed gray lines). The horizontal gray line emphases the zero value to guide the eye. Each data point shown is computed from ${\approx}5$ satellites per realisation that contributes. The median disruption times are essentially the same, but the distribution shows large scatter (the error bars show the range between $16$th and $84$th percentile). Intermediate mass satellites might survive marginally ($\sim 250\,\mathrm{Myr}$) longer in the simulations with higher dark matter resolution (grey). On the other hand, they seem to get destroyed marginally faster in the simulation with $64\times$ better stellar mass resolution (orange).}
    \label{fig:satellitedisruption}
\end{figure}

\section{Properties of satellite galaxies with the Superstars method}
\label{sec:satellites}

In addition to exploring the properties of the central galaxy, it is also important to look at those of the satellite population, which has been analysed already for the Auriga simulations with a focus on the satellites themselves \citep{Simpson2018, Grand2021} and their influence on the central galaxy \citep{Gomez2016,Gomez2016b,Simpson2019} and the stellar halo \citep{Monachesi2016,Monachesi2019,Riley2024,Shipp2024,Vera2025}. We want to ensure that properties of satellites remain unchanged with the \superstars\ method, not only to analyse the satellites themselves, but also because they set the properties of the Milky Way-mass host's accreted component. Their debris is critical for the formation of the stellar halo, as well as for substructure in the stellar disc and stellar halo. Most of the satellites are also significantly less massive than the central galaxy, so the numerical effects of the \superstars\ method in this different mass range may be different. Comparing the properties of satellite galaxies is therefore an independent check of the method.

Figure~\ref{fig:satellites} shows the satellite stellar mass function for all satellites within $R_\mathrm{200c}$ of the main galaxy which have a stellar mass greater than (or equal to) that of a single star particle in the reference simulations. The satellite mass functions are completely consistent between all simulations shown for a stellar mass greater than $10^6\,\mathrm{M_\odot}$. This mass is equivalent to $20$ star particles in the reference simulations. There seems to be a small systematic shift to fewer satellites for the lowest mass satellites with $M_*<10^6\,\mathrm{M_\odot}$ at higher stellar mass resolution. In contrast, when we increase the dark matter mass resolution only, the number of satellites with stellar masses smaller than $10^6\,\mathrm{M_\odot}$ increases by a similar small amount. The latter is likely the same effect as shown in \citet{Grand2021}, where low-mass dark matter halos are better resolved and form in higher numbers.

In Figure~\ref{fig:satellitesmassradius} we compare the sizes of satellite galaxies at $z=0$ for the different simulations. The sizes are fully consistent with each other. In the lowest mass satellite galaxies in the simulations with better dark matter mass resolution the scatter in sizes extends to lower values. This could be a result of numerical heating of the less resolved stellar particle population \citep{Ludlow2023} that could be relevant for poorly resolved low mass galaxies, even if it is irrelevant for the central galaxy of our main halo. This numerical heating goes away when the dark matter mass resolution is improved and the dark matter particles become slightly less massive than the star particles. For more massive satellite galaxies there are likely enough dark matter particles such that the artificial heating of the stars becomes irrelevant.

\begin{figure*}
    \centering
    \includegraphics[width=\textwidth]{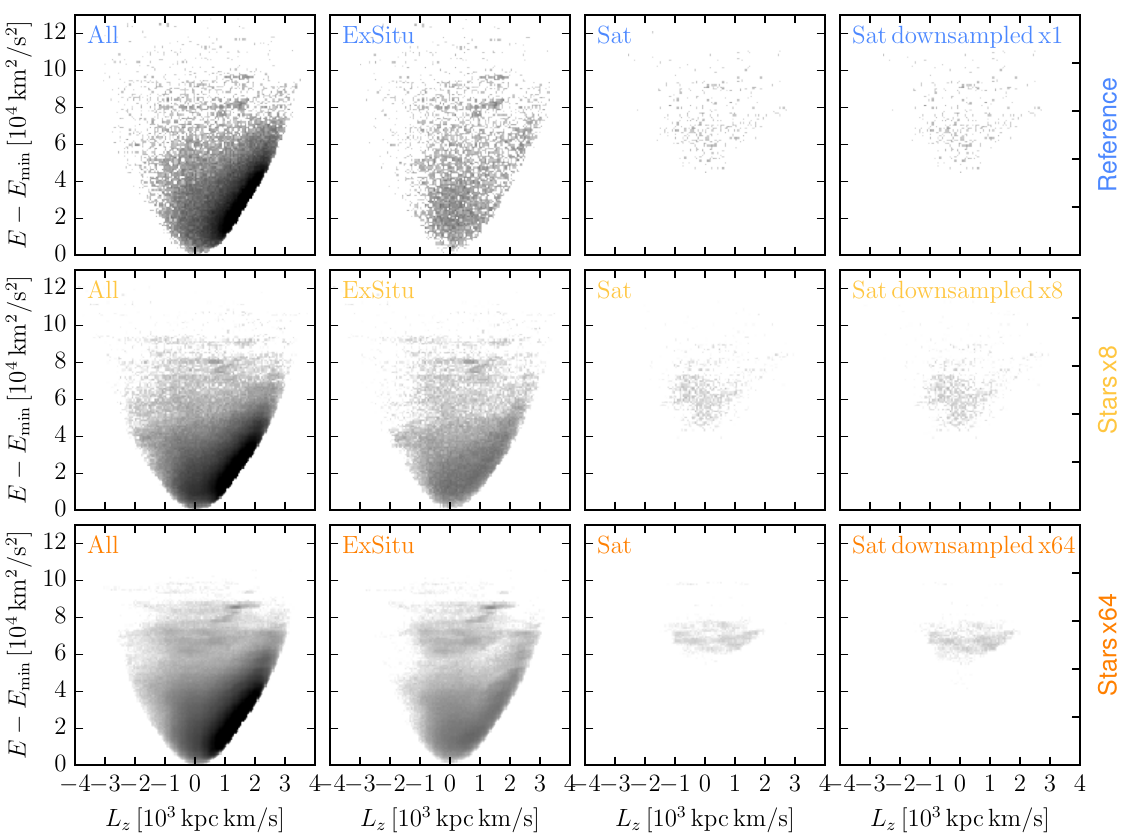}
    \caption{Stars that are in the galactic solar circle ($7\,\mathrm{kpc}<R_\mathrm{disc}<9\,\mathrm{kpc}$ and $\left|z\right|<10\,\mathrm{kpc}$) at $z=0$, shown in total energy--angular momentum space. The columns show \emph{all} stars (first column), accreted \emph{ex-situ} stars only (second column), the stars of one destroyed \emph{satellite} (third column), and the same satellite again but \emph{downsampled} to only showing one star per star formation event (fourth column). The rows show one reference run (first row), and one simulations with $8\times$ (middle row) and $64\times$ (bottom row) better stellar mass resolution. In the displayed 2D histogram, darker greys correspond to more stars in the bin. To compute the total energy shown on the y-axis, we subtracted the minimum total energy from all energies to remove different reference points in the potential. The substructure in the first two columns becomes much more clearly visible with better stellar mass resolution, while staying qualitatively similar. The stars from the single satellite shown in the last two columns are more coherent in energy--angular momentum space at better stellar mass resolution. In particular, this difference remains also when we downsample by the same factor as we used when increasing the stellar mass resolution.}
    \label{fig:substructure}
\end{figure*}

Having compared satellites that survive until $z=0$, we now compare the times when satellite galaxies are destroyed. In Figure~\ref{fig:satellitedisruption} we compare the satellite destruction times relative to the reference runs. To obtain the timings we first match the destroyed satellites in the different simulations via the Lagrangian regions of their $100$ most bound dark matter particles at the time when they reach their maximum stellar mass \citep[see also,][]{Riley2024}. We then compute the median of the destruction times in the reference runs for each satellite, and use this as a reference. We define the destruction time of a satellite as the time when its stellar mass first drops below $0.01$ of its peak stellar mass. We then compute, for all individual simulations, the destruction time of each satellite relative to the reference for this satellite. We then bin the satellites of all realisations of each simulation type in three stellar mass bins spanning a decade each in stellar mass. 

We can see that the destruction times are all also completely consistent with each other, albeit with large scatter. The only noticeable difference is in the intermediate stellar mass bin. Satellite galaxies with a peak stellar mass between $10^7\,\mathrm{M_\odot}$ and $10^8\,\mathrm{M_\odot}$ seem to survive typically $250\,\mathrm{Myr}$ longer in simulations with $8\times$ better dark matter mass resolution. Similarly, they might be destroyed slightly earlier in the simulations with $64\times$ better stellar mass resolution, but more realisations of this setup would be needed to understand if this difference is systematic or because of variance. We conclude that the properties of satellite galaxies and their destruction also remain unchanged when we increase the stellar mass resolution or the dark matter mass resolution. This makes the \superstars\ method an attractive path to study substructure from destroyed satellites galaxies in detail.

\begin{figure*}
    \centering
    \includegraphics[width=\textwidth]{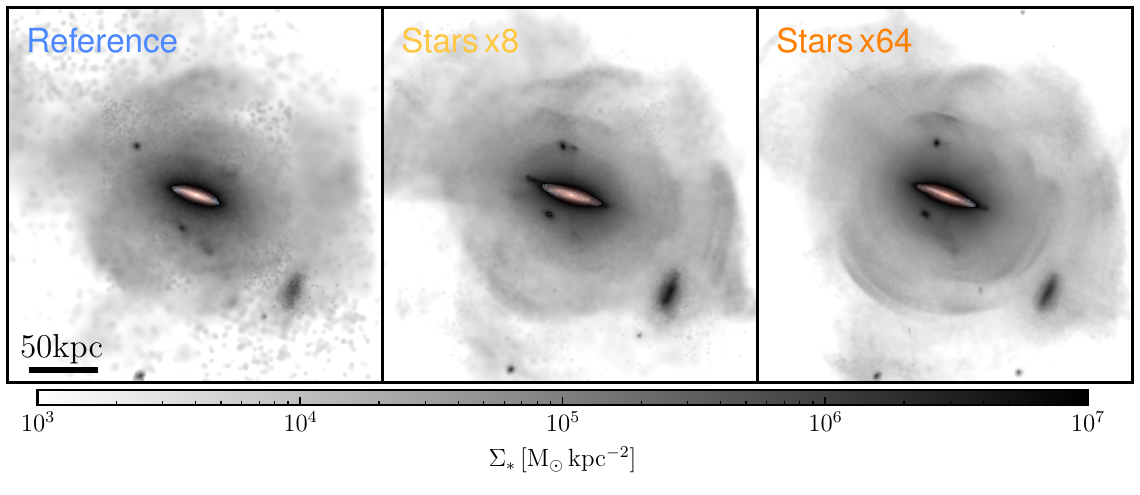}
    \caption{Projected stellar surface density of the galactic halo for the reference run (left panel), and simulations with $8\times$ (middle panel) and $64\times$ (right panel) better stellar mass resolution. For stellar surface densities larger than $10^7\,\mathrm{M_\odot\,kpc^{-2}}$ the panels show stellar light projections instead, analogously to Figure \ref{fig:stellarlight}, to indicate the location of the galactic disc within the halo. Better stellar mass resolution makes features in the stellar halo more visible and sharper, while remaining qualitatively very similar.}
    \label{fig:stellarhalo}
\end{figure*}

\section{Satellite destruction and debris with the Superstars method}
\label{sec:debris}

Having established that the \superstars\ method retains the properties of galaxies in our zoom simulations of Milky Way-like galaxies, and in particular of the stellar disc as well as the properties of satellite galaxies, we now focus on analysing the debris of destroyed satellites. We aim to understand whether the \superstars\ method helps modelling and understanding this debris both as substructure in the stellar disc and as part of the stellar halo.

We first look at substructure in the stellar disc and inner halo around the solar circle in Figure~\ref{fig:substructure}. We show mass-weighted histograms with a logarithmic colourmap of stars in energy -- angular momentum space, a classic method to find substructures in the Milky Way disc around the Sun \citep{Helmi2000,Gomez2017}. The rows show one of the reference simulations (top row), and two simulations with the \superstars\ method, one with $8\times$ better stellar mass resolution (middle row), and one with $64\times$ better stellar mass resolution (bottom row). We show star particles in the galactic solar circle, that is within $7\,\mathrm{kpc} < R_\mathrm{cyl} < 9\,\mathrm{kpc}$ and $\left|z\right| < 10\,\mathrm{kpc}$ at the present day. The first column shows a histogram of all star particles, the second column shows only star particles that are classified as ex-situ \citep{Grand2024}, that is, star particles that were originally formed in satellite galaxies. The third column shows only stars that originate from a single satellite galaxy, chosen to be still relatively coherent in specific total energy (sum of specific kinetic and potential energy) -- angular momentum space, matched between the simulations. The fourth column shows the same as the third column, but downsampled by a factor of $8$ (middle row) and $64$ (bottom row). We do this by selecting exactly one star particle per star formation event but counting it with the mass of all particles born at this event. Therefore, the number of star particles that contribute to each histogram in the right-most column is the same for all three simulations.

In the first column we already see that substructure becomes more visible and sharper from the top panel to the bottom panel with improved stellar mass resolution. This effect becomes even more clear when we only show ex-situ star particles in the second column. Clear horizontal substructures, that are structures with constant energy, emerge. In the third column we show this effect more explicitly for stars originating from one individual satellite. The stars of this satellite occupy a similar range in angular momentum in all three simulations, independent of stellar mass resolution. However, the spread in energy shrinks significantly with better stellar mass resolution. This makes the substructure appear much more distinctly in energy -- angular momentum phase space. One reason is that with better stellar mass resolution the stellar population of the satellite is sampled better, reducing the Poisson noise in the distribution and enhancing the visibility of features in the distribution. 

Additionally, better stellar mass resolution also reduces numerical noise in the gravitational potential, in particular in the stellar disc where the potential is dominated by the stars. This could reduce mixing in phase space due to numerical noise. To quantify this second, more subtle effect we downsample the number of stars in the simulations with better stellar mass resolution in the fourth column of Figure~\ref{fig:substructure}. Now, even though a similar number of star particles contribute to the panels in the right-most column, the features in the substructure are still more clear in the simulations with better stellar mass resolution. We argue that this demonstrates an improved, more faithful, modelling of stellar substructure beyond just better sampling, but also integrating of stellar substructure (and stars in general) in a more accurate galactic potential. Interestingly, the spread in energy keeps shrinking from the simulation with $8\times$ better stellar mass resolution to the simulation with $64\times$ better mass resolution, so even the latter might not be converged yet. Future work will need to quantify exactly what stellar mass resolution is required to obtain converged results. This will likely not only depend on the mass of the infalling satellite, but also on its orbit.

Finally, we look at the stellar halo and how it changes with better stellar mass resolution in Figure~\ref{fig:stellarhalo}. We show the stellar mass surface density at $z=0$ for one of the reference simulations, and simulations with $8\times$ and $64\times$ better stellar mass resolution with the \superstars\ method. To obtain the surface density maps we first calculate, for every star particle, the radius that encloses exactly $48$ star particles. We then use this radius to project the mass of the star particle with a 2D spline kernel onto a Cartesian grid. We include all stars within the halo, that is, within a radius of $R_\mathrm{200c}$. In the centre, for stellar mass surface densities larger than $10^7\,\mathrm{M_\odot\,kpc^{-2}}$, we show stellar light projections similar to Figure~\ref{fig:stellarlight} instead. Each projection is along the $z$-axis of the parent box which allows us to compare the orientations of the discs as well as the positions and orientations of satellites and other substructures in the stellar halo.
The stellar halo looks qualitatively similar between the simulations with different stellar mass resolutions, and the orientation of the stellar disc as well as most substructures are well matched between the simulations. We also clearly see that with improved stellar mass resolution features in the stellar halo, in particular shell structures and stellar streams, become sharper and better visible. This is likely mostly an effect of improved sampling, because the dynamics in the halo is dominated by the dark matter potential.

\section{Summary and Outlook}
\label{sec:summary}

We introduced the \superstars\ method which increases the stellar mass resolution in cosmological galaxy simulations, without increasing the gas mass resolution and with negligible additional computational cost. We showed in Section~\ref{sec:global} that increasing the stellar mass resolution with the \superstars\ method does not change the global properties or the dynamical state of a Milky Way-mass galaxy or its satellite galaxies in a full cosmological zoom simulation. In particular, the stellar mass distribution in the disc and the radial and vertical velocity dispersions of the stars do not change with the \superstars\ method in most of the disc (see Figure~\ref{fig:stars}). They show an interesting trend to slightly colder stellar discs with better resolutions in the outer parts of the disc at $R_\mathrm{disc}\gtrsim20\,\mathrm{kpc}$ that will require further study. 

We quantified various properties of the disc at $z=0$ and summarise them in Table~\ref{tab:simulations}. We then showed that the properties of satellite galaxies, their number counts (see Figure~\ref{fig:satellites}), their radii (see Figure~\ref{fig:satellitesmassradius}), and the timing when satellites get destroyed (see Figure~\ref{fig:satellitedisruption}) also do not change when we improve the stellar mass resolution with \superstars.
We finally looked at substructure in the stellar disc and the inner halo in Figure~\ref{fig:substructure} and in the stellar halo in Figure~\ref{fig:stellarhalo}. We showed that the \superstars\ method significantly improves the visibility and sharpness of substructure features in both the stellar disc and the stellar halo. We also demonstrated in Figure~\ref{fig:substructure} that this improvement is a combination of better sampling of the stellar distribution, as well as reduced noise in the stellar potential that leads to a more accurate integration of stellar trajectories, in particular in the outskirts of the disc and for substructures.

The optimal choice for the additional velocity kicks we impart on newly born stars in the \superstars\ method is not obvious, and can be seen as part of the subgrid model of star formation. When we initially used the \superstars\ method, we chose slightly larger kick velocities ($\sim 10\,\mathrm{km/s}$ in \citet{Grand2023}, about twice the value that we settled on and used in this work, i.e. $\sim\,5\,\mathrm{km/s}$). In Figure~\ref{fig:stars_kicks} in Appendix~\ref{sec:app} we show that this choice is unimportant for most galaxy properties. Still, it might be worthwhile to experiment more in future work to find an optimal choice.

Because increasing the stellar mass resolution with the \superstars\ method is significantly computationally cheaper than increasing the mass resolution of all components (gas, dark matter, stars), this method will allow us to run cosmological galaxy simulations with an unprecedented stellar mass resolution and more faithful stellar dynamics. This will allow us to better understand and interpret the new and upcoming wealth of Milky Way data, and help us better understand the physics of dynamical features like stellar bars, spiral arms, stellar substructure, and stellar streams in a realistic cosmological environment.

The \superstars\ method is useful for simulations in which star particles represent average stellar populations, but not when they represent individual stars. Simulations that achieve sufficiently high gas mass and stellar mass resolution to resolve individual stars are complementary to the \superstars\ approach. These simulations treat stars more realistically, however, computational cost limits them to  dwarf galaxies \citep{Hu2016, Wheeler2019, Agertz2020, Lahen2020, Gutcke2021, Gutcke2022a, Gutcke2022b, Deng2024} in the foreseeable future. The \superstars\ method will work best, that is preserve the properties of the galaxies while improving the stellar mass resolution in a controlled way, only when it does not touch the feedback model. That is in particular the case for feedback models that directly model galactic winds \citep[e.g.][]{Vogelsberger2014Illustris,Auriga,TNGMethodsPillepich,Smith2024Arkenstone}. In contrast, if star particles impart feedback locally  on the gas in discrete events (via thermal or kinetic energy injection), and star particles cause more than one feedback event during their lifetime \citep[e.g.,][]{Smith2018, FIRE2, Marinacci2019, Bieri2023}, using the \superstars\ model can change the spatial correlations between feedback events.

It is interesting to look at the computational cost of going to better stellar mass resolution. Note that these values are just a general guideline and not a precise estimate, because most of the simulations were run on different machines and with different code versions (in particular the original Auriga simulations). As a baseline, in the default setup the most costly part is evolving the gas. When improving the mass resolution of all components (but most critically the gas) by a factor of $8$, the computational cost typically increases by about a factor of $20$, due to $8\times$ more resolution elements, twice the number of timesteps, and some additional parallelisation losses. In contrast, improving the stellar mass resolution by a factor of $8$ with the \superstars\ method only increases the computational cost by $\sim10\%$ for our simulations, as star particles are a subdominant contribution to the total cost. Improving the stellar mass resolution by a factor of $64$, however, increases the total cost roughly by a factor of $6$. This is because now there are significantly more star particles in the halo at $z=0$ than dark matter particles and evolving the star particles becomes the dominant cost of the simulation. Improving the dark matter mass resolution by a factor of $8$ and keeping stellar mass and gas mass resolution the same increases the cost of the simulation roughly by a factor of $3$.

Finally, note that the main strength of the \superstars\ method is that it improves the stellar component. It will not help us to better understand structures in the interstellar medium. Notably, even though the sources of chemical enrichment (the star particles) are better resolved, metals will still be averaged to the gas resolution when they are returned to the gas from stars. Resolving the interstellar medium, or similarly the circumgalactic medium, or the chemical evolution of galaxies better will therefore still require improving on the gas resolution, with all practical problems that come along with it. In principle, it should be possible to improve or redesign the models for gas physics such that galaxy properties become independent of gas resolution, but it will require significant future work. Therefore, the \superstars\ method seems to be a better approach to improve the stellar mass resolution in cosmological galaxy simulations for the near future.

\section*{Data availability}
The data underlying this article will be shared on reasonable request to the corresponding author. 

\section*{Acknowledgements}

RP thanks Anna Genina for helpful discussions. 
RB is supported by the UZH Postdoc Grant, grant no. FK-23116 and the SNSF through the Ambizione Grant PZ00P2\_223532.
RJJG acknowledges support from an STFC Ernest Rutherford Fellowship (ST/W003643/1).
FvdV is supported by a Royal Society University Research Fellowship (URF\textbackslash R1\textbackslash191703 and URF\textbackslash R\textbackslash241005).
FAG acknowledges support from the ANID BASAL project FB210003, from the ANID FONDECYT Regular grants 1251493 and from the HORIZON-MSCA-2021-SE-01 Research and Innovation Programme under the Marie Sklodowska-Curie grant agreement number 101086388. FF is supported by a UKRI Future Leaders Fellowship (grant no. MR/X033740/1).
The authors gratefully acknowledge the Gauss Centre for Supercomputing e.V. (www.gauss-centre.eu) for compute time on the GCS Supercomputer
SUPERMUC-NG at Leibniz Supercomputing Centre (www.lrz.de) and the DiRAC@Durham facility which is managed by the Institute for Computational Cosmology on behalf of the STFC DiRAC HPC Facility (www.dirac.ac.uk). The equipment was funded by BEIS capital funding via STFC capital grants ST/K00042X/1, ST/P002293/1, ST/R002371/1 and ST/S002502/1, Durham University and STFC operations grant ST/R000832/1. DiRAC is part of the National e-Infrastructure.



\bibliographystyle{mnras}

\appendix

\section{Changing the kick velocity in the Superstars model}
\label{sec:app}

\begin{figure*}
    \centering
    \includegraphics[width=\textwidth]{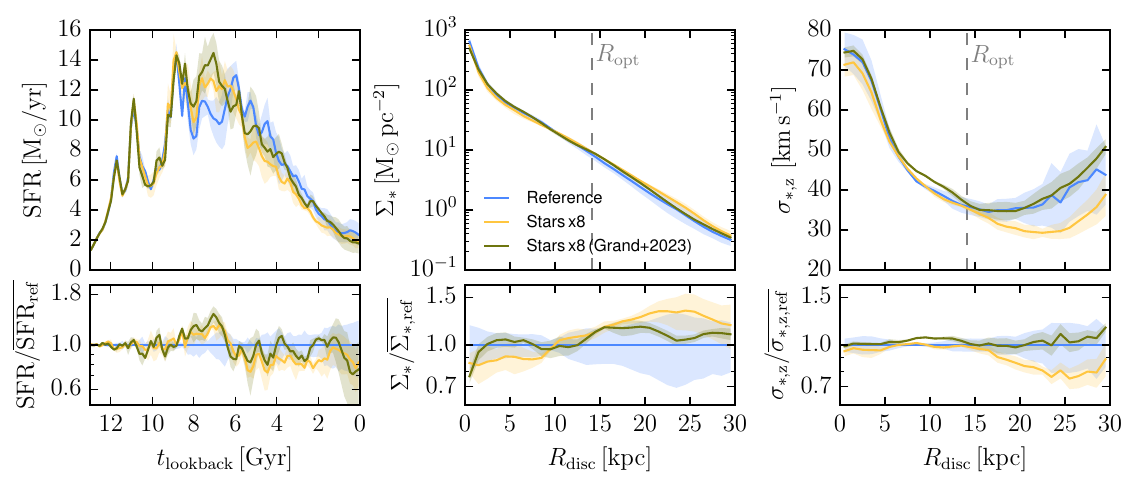}
    \caption{Comparison of the star formation history (left panel) and profiles of stellar mass surface density (middle panel) and vertical velocity dispersion (right panel) at $z=0$ for the reference simulations and two sets of \superstars\ simulations with $8\times$ better mass resolution and different initial velocity kicks. For consistency with the old simulations we show only $3$ realisations for each type. The lower row of panels shows the deviation relative to the mean of the reference simulations. The blue and yellow lines are the same as in Figure \ref{fig:stars}. The olive lines show a set of $3$ simulations using an older version of the superstars method \citep{Grand2023} with about two times larger kicks for the star particles at formation. The star formation history, surface density profile, and vertical velocity dispersion out to the optical radius are consistent between all sets of simulations. Only the vertical velocity dispersion outside the optical radius changes slightly with the \superstars\ method and different kick velocities (see also discussion in Sec.~\ref{sec:global}).
    }
    \label{fig:stars_kicks}
\end{figure*}

In Figure~\ref{fig:stars_kicks} we compare the \superstars\ method as described in Section~\ref{sec:method} with an earlier version that uses slightly larger kick velocities \citep{Grand2023}. We find that the star formation histories and stellar surface density profiles are consistent between the reference Auriga simulations and both versions of the \superstars\ method. Only the outer parts of the stellar disc, outside the optical radius, become slightly hotter with larger kicks \citep[as used in][]{Grand2023} and slightly colder with the version we present here, compared to the reference Auriga simulations without improved stellar mass resolution.

% Don't change these lines
\bsp	% typesetting comment
\label{lastpage}
\end{document}